# Design and Build of a Customizable Light-Sheet Microscope for Visualization of Cytoskeleton Networks

Nathan Felcher[1], Daisy Achiriloaie[1], Brian Lee[2], Ryan McGorty[3], Janet Sheung[1,2,4]

[1]W.M. Keck Science Department, Scripps College, Claremont, CA, United States
[2]W.M. Keck Science Department, Claremont McKenna College, Claremont, CA, United States
[3]Department of Physics and Biophysics, University of San Diego, San Diego, CA, United States
[4]W.M. Keck Science Department, Pitzer College, Claremont, CA, United States

**CORRESPONDING AUTHOR:** Janet Sheung (jsheung@kecksci.claremont.edu)

**INTRODUCTION:**

Light-sheet microscopy (LSM) refers to a family of high-resolution fluorescence imaging techniques where the excitation light is shaped into a sheet[1, 2], including Selective Plane Illumination Microscopy (SPIM), Swept confocally-aligned planar excitation (SCAPE), and oblique-plane microscopy (OPM)[3–7]. Compared to better-known microscopy modalities such as epi-fluorescence, Total-Internal Reflection Fluorescence Microscopy (TIRFM), or confocal, in LSM only the plane of the sample being actively imaged is illuminated, thereby minimizing phototoxicity and lengthening the timescales over which samples can be imaged[8–10]. This gentle optical sectioning capability allows LSM techniques to excel at imaging three-dimensional samples over extended time periods, notably even those too thick for confocal microscopy techniques. For these reasons and more, since its original development in 2004 LSM has become the imaging technique of choice for many physiologists, developmental biologists, and neuroscientists seeking to visualize entire organisms such as live zebrafish and drosophila embryos[3, 4, 6, 11]. In the two decades since, the advantages of LSM have been leveraged to visualize structure and dynamics at progressively smaller scales: tissue[11, 12], cellular and subcellular both in vivo and in vitro[13–17].

Despite this preponderance of successful use-cases in the literature, the high cost of commercial LSM systems (~ a quarter million USD as of this writing)[18, 19] remains a hurdle to more widespread use of the technique. To make DIY builds a tractable alternative for researchers, multiple build guides have been published[8, 13, 20, 21], including the open-access effort OpenSPIM[22]. However, to date, detailed build guides accessible to researchers with minimal optics experience are limited to earlier LSM designs which are incompatible with traditional slide-mounted samples **(Figure 1A).** In contrast, recent single-objective light-sheet (SOLS) implementations circumvent this limitation by utilizing a single objective for both excitation and detection **(Figure 1C)**[5, 6, 8, 13, 20]. The cost for the versatility of the SOLS design is a substantial increase in complexity of the build, as two additional objectives are required to relay then de-tilt and reimage the object plane onto the camera for imaging **(Figure 1D).** To make the complex SOLS-style setups more accessible, this article and accompanying video protocol presents a reasonably self-contained, step-by-step guide on the design, build, alignment process, and use of a slide-compatible SOLS system assuming prior knowledge only at the level of an entry level optics course.

Although the protocol itself is succinct, we direct readers to other resources in the preparation steps to learn more about particular parts of the design or hardware considerations. If a reader intends to follow the specifications of this design, it may not be necessary to understand how to select particular optical components. If a user wishes to modify the system for their needs, however, these resources provide crucial information on topics such as hardware compatibility and important details about changing the design.

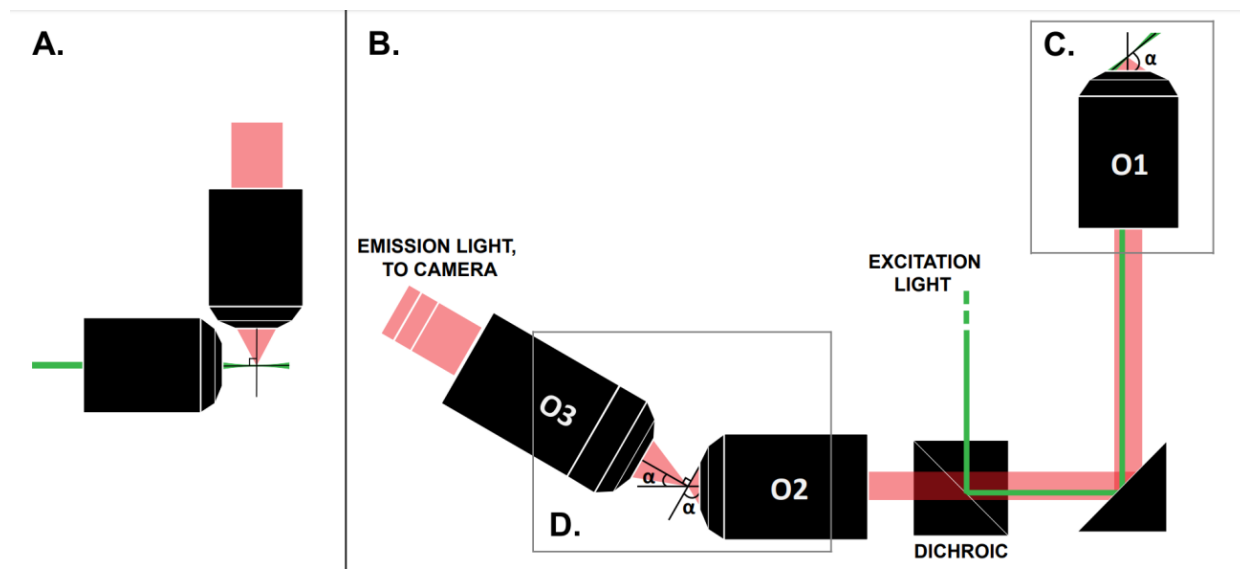

**Figure 1: Characteristics of different LSM configurations. A.** The two orthogonal objectives setup common in early LSM designs. This configuration is not compatible with traditional slide mounting techniques, and instead uses a capillary tube or a cylinder of gel to contain the sample. **B.** An abbreviated schematic of a SOLS light-sheet design, highlighting: **C.** The use of a single objective for both excitation and emission collection at the sample plane (O1), allowing for a traditional slide to be mounted on top and **D.** The relay objective system in the SOLS emission path. O2 collects the emission light and de-magnifies the image. O3 images the plane at the correct tilt angle onto the camera sensor.

**PROTOCOL:**

1. **Preparation for alignment:**

1.1) Before starting the build, perform any necessary literature reviews to have a clear idea of the intended use case (e.g. fluorophores to be imaged, necessary imaging volume, resolution requirements, etc.). In particular, refer to the representative results section below to decide whether following the exact design described here is appropriate. If yes, skip to step 1.2. If no, find suggestions and guidance for hardware selection at the Sheunglab SOLS Build Guide[23].

1.2) Collect all of the necessary optical, opto-mechanical and electrical components as detailed in the Table of Materials. For users modifying the system, collect all equivalent parts.

1.3) Build the alignment laser as depicted in **Figure 2A.** Check that the beam is collimated using the shear plate.

1.4) Build the double frosted glass disk alignment cage as depicted in **Figure 2B.**

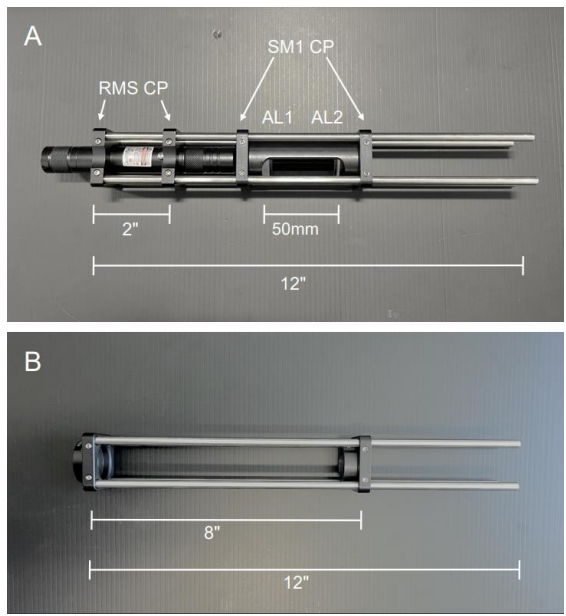

**Figure 2: Photos of alignment tools. A.** Collimated alignment laser. RMS CP: RMS threaded cage plate; SM1 CP: SM1 threaded cage plate; AL1: Alignment lens 1, -50mm; AL2: Alignment lens 2, 100mm **B.** Double frosted glass disk alignment cage.

1.5) Prepare the fluorescent dye coated test sample

1.5.1) Create a saturated rhodamine dye solution by gradually adding water to a small amount of lyophilized powder until all are dissolved. Vortex to homogenize.

CAUTION: Always wear gloves and a mask when handling rhodamine powder.

1.5.2) Pipette 10µL onto the center of a test slide.

1.5.3) Place a 22mm x 22mm cover glass on top of the liquid, ensuring that the layer of fluorescence is as thin as possible.

1.5.4) Seal with clear nail polish.

1.6) Prepare the 3D bead sample (1µm beads embedded in gel)

1.6.1) Use double sided tape to create a 4-5mm wide vertical channel on a sample slide three layers of tape high.

1.6.2) Place a 22mm x 22mm cover glass on top of the tape in the center of the slide. Press firmly on the taped regions to ensure a good seal between the tape and the cover glass.

1.6.3) Use a razor blade to remove the excess tape.

1.6.4) Prepare a saturated gelrite solution by gradually adding gelrite powder into DI water. This solution will be solid at room temperature and liquid at 65° C.

1.6.5) Prepare 10µL of 1:1000 dilution of beads in warmed gelrite solution.

1.6.6) Carefully pipette the solution into the chamber until full and seal the edges with epoxy.

2. **Align the excitation path:**

2.1) Sketch out the microscope layout on the surface of the optical table. Measure all distances as accurately as possible.

NOTE: Refer to **Figure 3** for the location of components within the system.

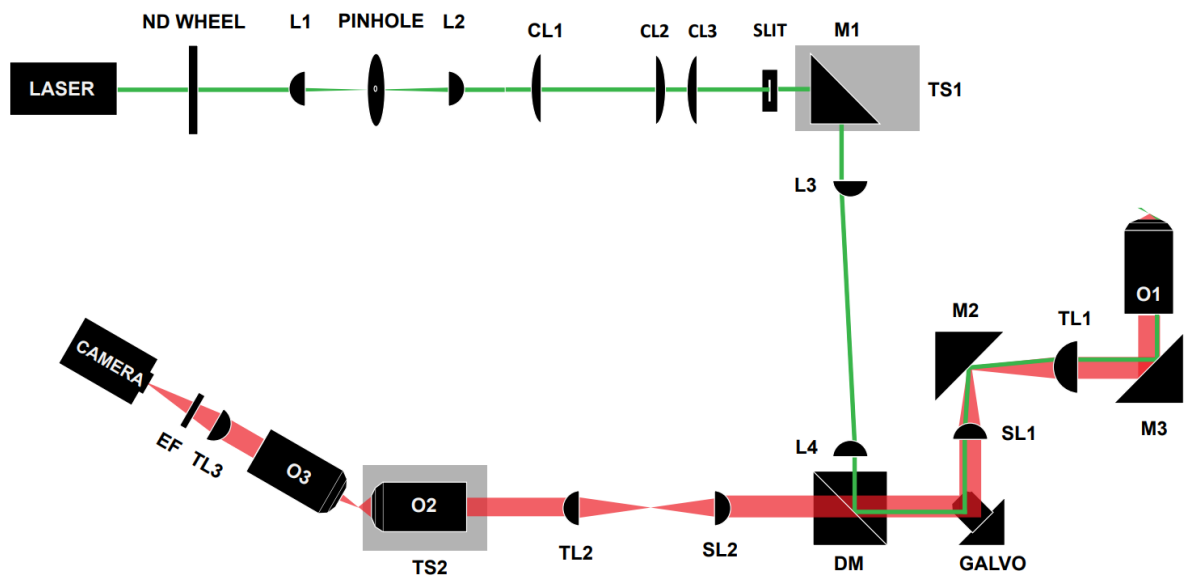

**Figure 3:** Schematic layout of SOLS system with all components labeled. Excitation path shown in green. Emission path shown in red.

ND Wheel: Variable neutral density filter wheel; L1-L4: Plano concave achromat lenses; CL1-CL3: Cylindrical lenses; M1-M3: Mirrors; TS1-TS2: Translation stages; DM: Dichroic mirror; Galvo: Scanning galvanometer; SL1-SL2: Scan lenses; TL1-TL2: Tube lenses; O1-O3: Objectives. EF: Emission filter. Focal lengths of lenses: L1:100mm; L2:45mm; CL1: 50mm; CL2: 200mm; CL3: 100mm; L3:150mm; L4: 100mm; SL1: 75mm; TL1: 200mm; SL2:150mm TL2: 125mm TL3: 200mm See parts table for more detailed part specifications.

2.2) Mount the excitation laser onto the table. Set two irises to the intended height of the laser and mount them 2-3 feet apart on the desired line of holes behind the location of M1. Use these irises to ensure that the beam is level to the table surface and centered on the line of holes on the optical table.

NOTE: Wear laser safety glasses for eye protection and block off any stray laser beams with laser safety screens as a safety precaution.

NOTE: Until all components are permanently clamped down, drift on the order of hours is possible. Set up an iris at the furthest point of the alignment at the end of each day as a quick visual check for drift when returning to the build.

NOTE: Vibrations, improperly floated optical table, and air currents are the most common causes of optical drift.

2.3) Mount the laser shutter as close as possible to the excitation laser. Utilize this shutter to quickly block the laser light during alignment, rather than repeatedly turning the laser on and off.

2.4) Assemble the ND filters into the ND filter wheel (ND 0.5, 1.0. 2.0, 3.0, 4.0 and a blank slot) and mount after the laser shutter.

2.5) Switch the motorized actuator onto one translation stage (TS1) and then insert the stage under the location of M1. The stage must translate axially along the same line of holes that the laser light follows. Set the stage in the middle of its range for the initial placement.

NOTE: In steps 2.6-2.10, insert reflective optical components into the beam path one at a time to direct the laser along the path as drawn on the table. Use the pair of irises set to the exact height to define the desired exit beam path and to guide the placement and alignment of each reflective element. For each element, adjust the height and position of the mount to ensure that the incoming beam hits the center. Then, rotate the base of the mount to direct the beam along the drawn-out beam path on the table so that it passes through both irises. Fine adjust the tilt of the outgoing beam with the knobs on the back of each mount.

NOTE: After each element is aligned to the correct height, add a slip-on collar to the post to secure the height.

2.6) Mount and align M1 on top of TS1.

2.7) Mount and align the dichroic onto the table.

2.8) Following manufacturer's instructions, connect the galvo to the power supply and function generator.

2.9) Mount galvo such that laser is incident on exact center of mirror. Align the galvo with the mirror tilt set to 0V (center of range).

2.10) Mount and align M2.

2.11) Insert the large cylindrical mirror into the cylindrical mirror mount. Use 1" cage rods to attach a 30mm-60mm cage adapter above M3. Use the knobs on the mirror mount to flatten the tilt of the M3 mirror for the initial placement.

NOTE: The cage rods should not extend beyond where the top of the objective will be as to not interfere with future mounting of samples.

2.12) Mount the double frosted glass alignment cage into the cage adapter above M3. **It is crucial to tighten the set screws on the cage adapter that hold the alignment cage in place each time it is used for alignment.** Mount M3 to the table and adjust the height and position until the beam is roughly centered on both frosted glass alignment disks. Clamp M3 to the table and use the knobs on the back of the mirror to fine adjust the beam.

NOTE: All reflective elements in the excitation path are now set and should not be touched.

2.13) Mount L1 onto the table. For all initial lens placements, use a screw on lens target to center the incoming beam on the front of the lens. Adjust the tilt and lateral position of L1 until the beam is centered on both of the frosted glass plates on the alignment cage above M3.

2.14) Mount L2 in its respective position to create a 4f system with L1. Move L2 axially to obtain a collimated beam, using the excitation laser and the shear plate to check collimation. Adjust the tilt and lateral position of L2 to center the beam on both of the frosted glass disks above M3.

2.15) Mount the pinhole into an xy translation mount. Mount this on top of 1-D translation stage to provide fine axial translation. Mount the pinhole and stage onto a post and post mount and place it at the shared focal point in between L1 and L2. Adjust the pinhole by hand until the laser light can be seen through the pinhole.

2.16) Mount the sensor of a power meter immediately after the pinhole. Adjust the xy position of the pinhole to obtain a TEM00 beam profile. Then, adjust the pinhole axially with the 1-D stage to find the location of maximum transmission.

2.17) Mount L4 onto the table in its position and adjust the mount to the correct height. Adjust L4 axially so that the excitation beam is focused on the surface of the galvo. Adjust the tilt and lateral position of L4 to center the beam on both frosted glass disks above M3.

2.18) Mount L3 onto the table in its position and adjust the mount to the correct height. Use the excitation laser and the shear plate to check collimation of L3 and L4. Adjust the tilt and lateral position of L3 to center the beam on both frosted glass disks above M3.

2.19) Temporarily remove L3. Mount SL1 onto the table and adjust the axial distance until it forms a collimated telescope with L4, as measured with the shear plate. Adjust the tilt and lateral position of SL1 to center the beam on both frosted glass disks above M3.

2.20) Reinsert L3. Mount TL1 and use the excitation beam and shear plate to check collimation of the SL1 and TL1. Adjust the tilt and lateral position of TL1 to center the beam on both frosted glass disks above M3.

2.21) Using an adapter ring, screw O1 into the cage plate above M3. Remove SL1 temporarily and let the beam hit the ceiling. Adjust the height (axial distance) of O1 on the cage system until the beam forms an Airy disk on the ceiling, then continue adjusting until the size of the disk is minimized.

2.22) With O1 in place, mount the sample stage in the appropriate position.

3. **Align the emission path:**

3.1) Setup the alignment laser

3.1.1) Mount the alignment laser beside excitation laser using a post under each of the alignment lens cage plates.

3.1.2) Remove O1 and reinsert the frosted glass alignment cage. Use one kinematic mirror mount and one drop down mirror to align the alignment beam to the excitation beam.

3.1.3) Use the power meter to maximize the signal of the alignment beam after the pinhole by fine adjusting the two mirrors. Ensure that the beam remains centered on the frosted glass alignment cage.

3.2) Remove the alignment cage and reinsert O1. Place the square mirror on the sample stage of O1 and adjust the mirror axially until the size of the beam profile is minimized after the dichroic.

3.3) Mount one iris in the emission path, far enough back to insert SL2, TL2 and O2 without interfering. Align this iris to the alignment laser. Mount a frosted glass disk at least 1' behind the iris, also aligned to the laser light.

3.4) Insert SL2 at the correct distance, measured from the galvo with a ruler. Adjust the tilt and lateral position of SL2 so that incoming alignment beam is centered on SL2 and the outgoing beam passes through the iris and frosted glass disk.

3.5) Insert TL2 at the correct distance, measured from SL2 with a ruler. Adjust the tilt and lateral position of TL2 so that incoming alignment beam is centered on TL2 and the outgoing beam passes through the iris and frosted glass disk.

3.6) Mount TS2 onto the table. Ensure that the stage translates along the optical axis of O2.

3.7) Screw O2 onto an xy translation mount. Connect a post under the xy mount to mount O2 onto the translation stage. Use a screw on target to center the back of O2 on the red laser.

3.8)  Adjust the xy knobs and tilt of O2 so that the red alignment beam passes through the iris and frosted glass disk.

3.9)  Move the alignment laser to above O1, directed downwards into the emission path. Turn on the laser and ensure that this beam is centered on all lenses in the emission path.

3.10) **Conjugate the pupil planes of O1 and O2:** Open the iris and remove the frosted glass disk so that the alignment beam exiting O2 continues unobstructed to a faraway surface or wall (>0.5m) **(Figure 4). Adjust TS2 until the beam forms a small Airy disk on the surface then continue adjusting TS2 to minimize the size of the Airy disk.**

   NOTE: A strongly diverging beam indicates incorrect placement of O2. However, because this beam passes through two objectives, a small amount of divergence is inherent. As such, the Airy disk is the best guide.

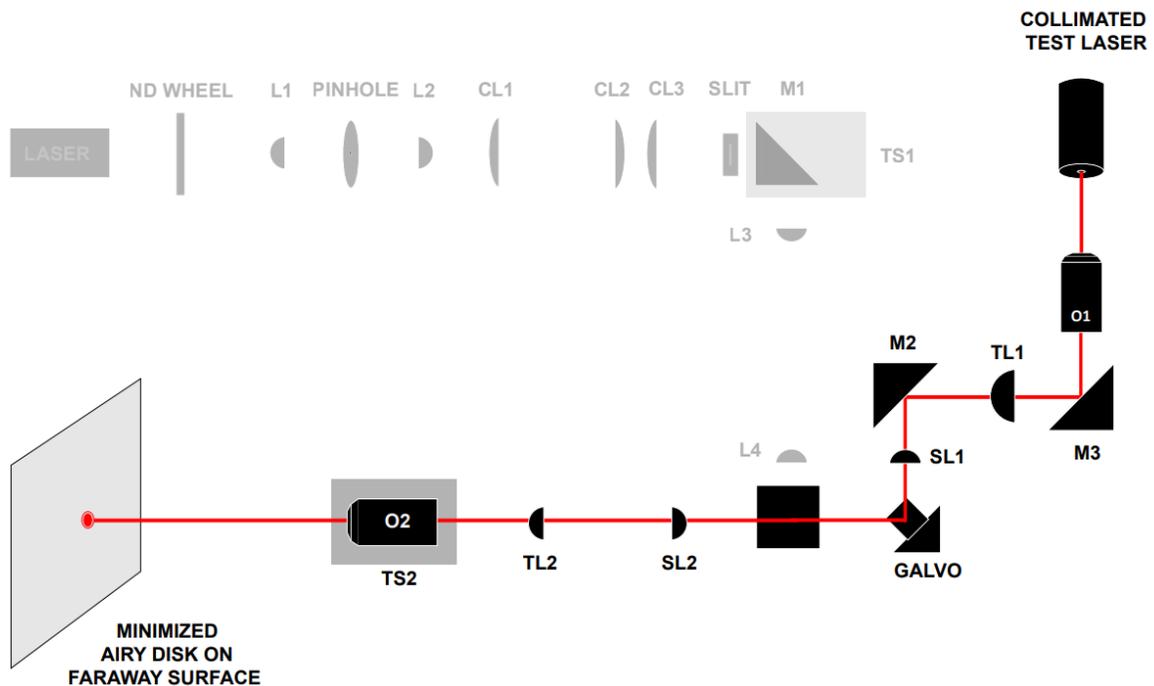

**Figure 4: Laser-in-laser-out technique**: sending a collimated test beam through the front of O1 and observing the beam that exits O2 on a faraway surface. If all components are aligned at the correct distance, the beam will form a small Airy disk on the faraway surface.

3.11) **Optimize the galvo for tilt-invariant scanning:** Power the galvo with a low frequency (~1 Hz) sawtooth signal and observe the alignment beam on the same faraway surface or wall. If the galvo is placed incorrectly, the beam will sweep horizontally on the surface along with the galvo movement. This can be resolved through fine adjustments (by hand) of the tilt and xy position of the galvo base until the beam displacement is undetectable by eye.

3.12) Check the quality of the system by imaging at 0°

3.12.1) Screw O3 into an xy translation mount. Screw a 1" lens tube into the cage translation stage and screw the xy translation stage into the tube. Use two cage rods to connect the front of the cage translation stage to a cage plate and mount the cage plate to a post. Mount O3 close to the front of O2 (~4-5mm) at 0° and adjust the height to match.

3.12.2) Mount a frosted glass alignment disk in the shared focal plane between SL2 and TL2, measured with a ruler. Mount the acrylic fluorescent test slide onto the stage and illuminate the slide with the excitation laser. Look through the back of O3 and adjust both the height and axial position of O3 to find the emission light, then adjust O3 axially until the emission light fills the back aperture **(Figure 5).**

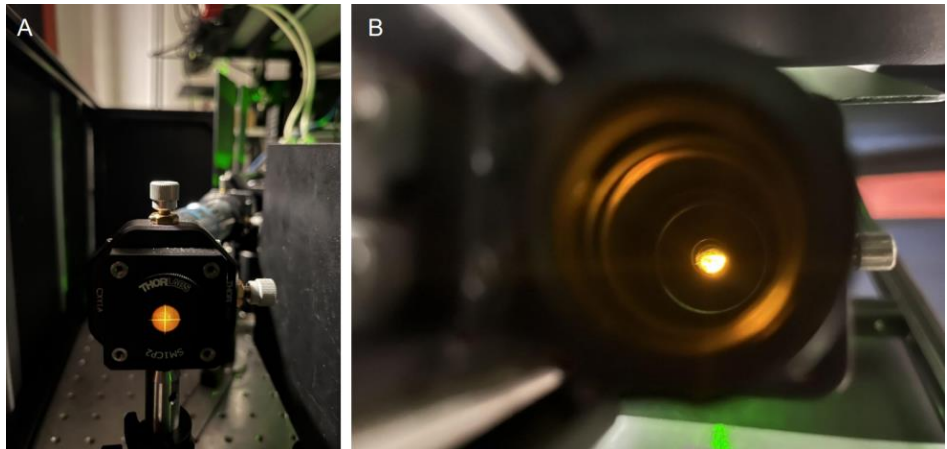

**Figure 5: Utilizing emission light for alignment. A.** Emission light from acrylic fluorescent slide on a screw-on target behind the BFP of O2. **B.** Finding the emission light by sight through the back of O3.

3.12.3) Screw in two 8" cage rods to the back of the O3 translation stage. Slide a cage plate with a mounted frosted glass disk onto the rods, then fine adjust O3 using the xy mount to ensure that the emission light exits O3 centered. Then, remove the cage rods.

3.12.4) Mount a frosted glass disk in the rough position of the camera sensor and align the height and position of the disk to the emission light.

3.12.5) Screw TL3 into a cage plate and mount it immediately behind O3. Center TL3 on the incoming emission light, then adjust the tilt of TL3 to align the outgoing light to the frosted glass disk.

3.12.6) Mount the camera at the correct distance from the tube lens, measured with a ruler.

3.12.7) Screw both 2" lens tubes and the emission filter onto the camera.

3.12.8) Remount the frosted glass alignment disk in the shared focal plane between SL2 and TL2. Mount the fluorescent dye test sample and illuminate the sample with the excitation beam.

3.12.9) Adjust the xy translation knobs on the O3 mount until the small hole in the glass alignment disk is centered within the field of view (FoV). Adjust O3 axially with the cage translation stage until the hole is in focus: the edges should appear sharp **(Figure 6).**

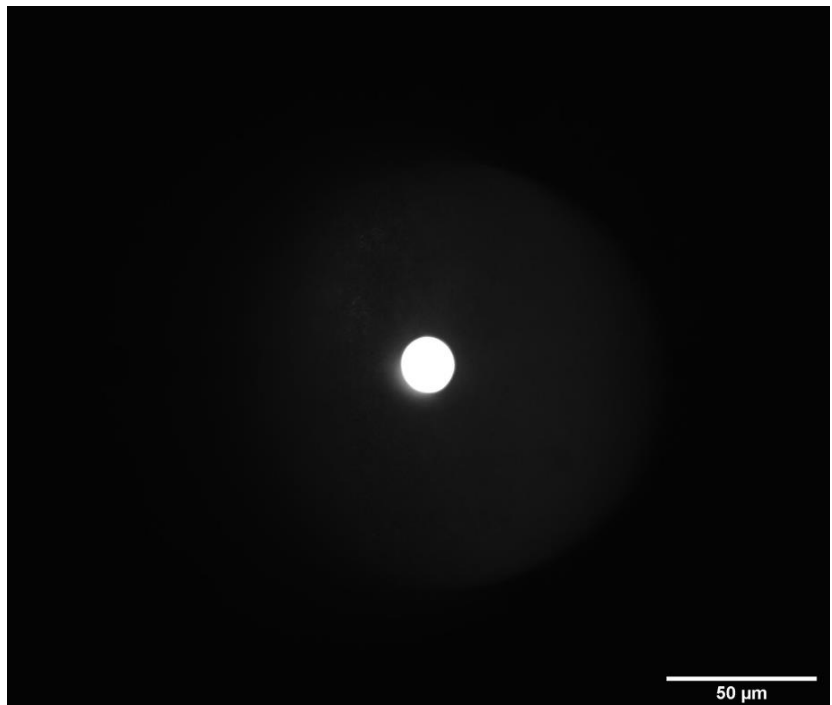

**Figure 6:** On camera image of the correctly focused frosted glass alignment disk placed at the intermediate plane between SL2 and TL2.

3.12.10) **Image fluorescent beads to check quality of the system:** Remove the frosted glass alignment disk. Mount the 3D bead sample and illuminate the sample with the excitation beam. Adjust the height of the sample relative to O1 until the fluorescent beads fill a circular region in the center of the FoV. Fine adjust the position of O3 using the xy stage and the axial translation stage until PSFs are round (indicative of minimal aberrations), and bright (indicative of good signal to noise ratio) **(Figure 7)**.

NOTE: If this cannot be achieved by adjusting O3, it is highly likely that the optical system in between components O1 and O2 is sub-optimally aligned, and

Page 11 of 27

the user should follow the diagnostic checks in the step below. **If round PSFs can be obtained, the user can skip the diagnostic step and move on to tilting the imaging system.**

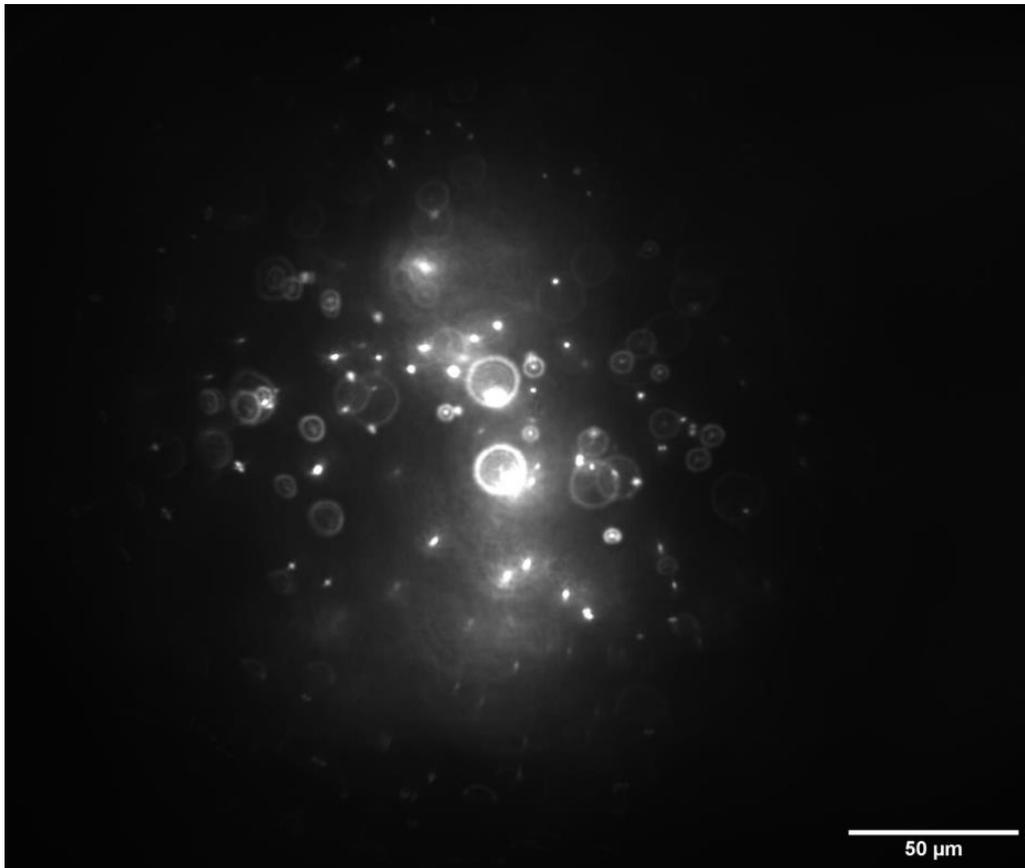

**Figure 7:** Camera image of the 3D bead sample (1nm beads) with the imaging module set to 0°, illuminated by a circular beam, prior to the insertion of the cylindrical lenses.

3.13) Perform diagnostic checks if needed

NOTE: Once good PSFs are obtained, the rest of the diagnostic steps can be skipped.

3.13.1) Mount the brightfield (BF) light above O1. Mount the positive grid test target on the sample stage **at the same axial height as the alignment mirror.** Center the 10 micron grid and illuminate the grid with the BF light.

3.13.2) Image the grid on camera and translate the sample until the grid is in focus. The grid image can be used to confirm that the field across the FoV is flat: if not, the grid will appear distorted and bowed. To correct for a bad grid image, adjust the xyz position and tilt of O3, then adjust the TL3 and the camera accordingly.

NOTE: If a flat grid can be achieved, repeat step 3.12.10 then move on to tilting the imaging system.

3.13.3) Set up the alignment camera or the imaging camera at the correct distance so that SL2 focuses the image onto the sensor. Image the grid target at the intermediate plane **(Figure 8)**. If this grid is also skewed, it is highly likely that the optical system in between components O1 and SL2 is sub-optimally aligned and should be re-visited. Optimize alignment as necessary before progressing.

NOTE: If the camera does not fit in between SL2 and TL2, use an extra mirror to bounce the image 90° after SL2 and onto the camera.

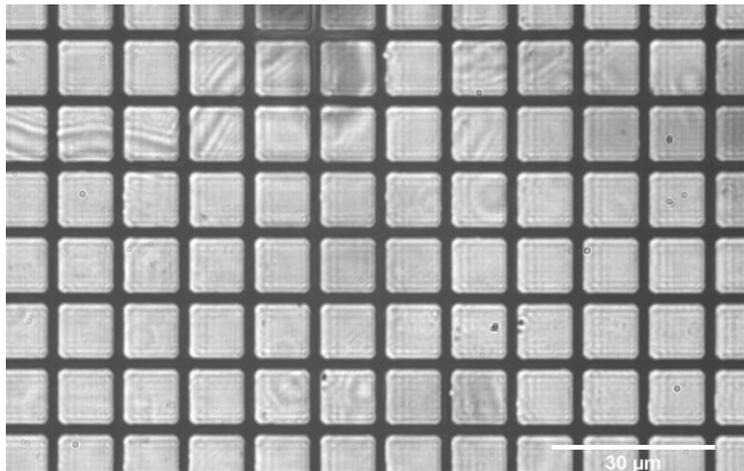

**Figure 8:** Positive grid test target correctly focused at the intermediate plane between SL2 and TL2. The flat grids throughout the entirety of the field indicate good alignment of components SL2 and prior.

3.13.4) Check the PSFs at the intermediate plane: After checking the grid, another good diagnostic check is to check the PSFs at the same intermediate plane. A good image at this plane, which is similar to Figure 7 but at a different magnification **(Figure 9),** indicates good alignment through SL2.

NOTE: If a flat grid and round PSFs can be achieved at the intermediate plane, repeat step 3.13.2, then step 3.12.10 and then move on to tilting the imaging system.

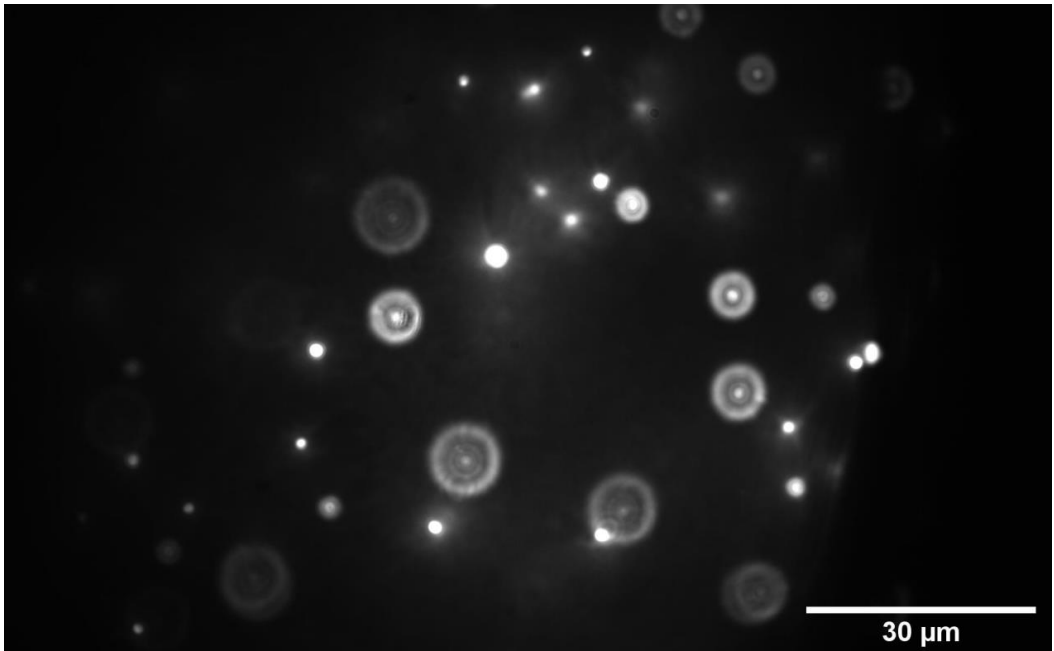
**Figure 9:** Camera image of the 3D bead sample (1nm beads) correctly focused at the intermediate plane between SL2 and TL2.

3.14)   Tilt the O3 imaging sub-system to 30°

3.14.1) Remove O3, TL3, and the camera.

3.14.2) Reinsert O3 at 30° to the optical axis of O2 using the lines on the table as a guide.

3.14.3) Mount a frosted glass alignment disk in the shared focal plane between SL2 and TL2. Mount the acrylic fluorescent test slide onto the stage and illuminate the slide with the excitation laser. Once again, look through the back of O3 and adjust both the height and axial position of O3 to find the emission light at 30°, then adjust O3 axially until the emission light fills the back aperture.

3.14.4) Remove the frosted glass alignment disk from in between SL2 and TL2 to obtain a stronger emission signal.

3.14.5) Screw in two 8" cage rods to the back of the O3 translation stage. Slide a cage plate with a mounted frosted glass disk onto the rods, then fine adjust O3 using the xy mount to ensure that the emission light exits O3 centered. Then, remove the cage rods.

3.14.6) Mount a frosted glass disk in the rough position of the camera sensor and align the height and position of the disk to the emission light.

3.14.7) Mount TL3 immediately behind O3. Center TL3 on the incoming emission light, then adjust the tilt of TL3 to align the outgoing light to the frosted glass disk.

3.14.8) To more accurately set the TL3-camera distance, carefully unscrew O3 and mount the alignment laser so that is focused by TL3 onto the camera. Use ND filters as necessary so laser intensity <1mW. Start camera live view and adjust TL3 axially to minimize the laser spot on camera.

3.14.9) Remount the frosted glass alignment disk in the shared focal plane between SL2 and TL2. Mount the fluorescent dye test sample and illuminate the sample with the excitation beam. Adjust the xy translation knobs on the O3 mount until the small hole in the glass alignment disk is within the FoV on camera. Adjust the cage translation stage to move O3 axially until the hole is in focus: the hole should appear similar to how it did at 0°.

3.14.10) Remount the positive grid test target at the same axial height and illuminate the grid with the BF light. Due to the 30° tilt, only one vertical section should be in focus. Once again, the grid image can be used to confirm that the field across the FoV is flat, even when out of focus. When the slide is translated axially, the in-focus portion of the FoV (grid target) should sweep across the screen horizontally while the grid squares maintain a consistent size **(Figure 10).**

   NOTE: Because of the tilt of the imaging plane at the sample, the grid may appear slightly stretched in the x-plane.

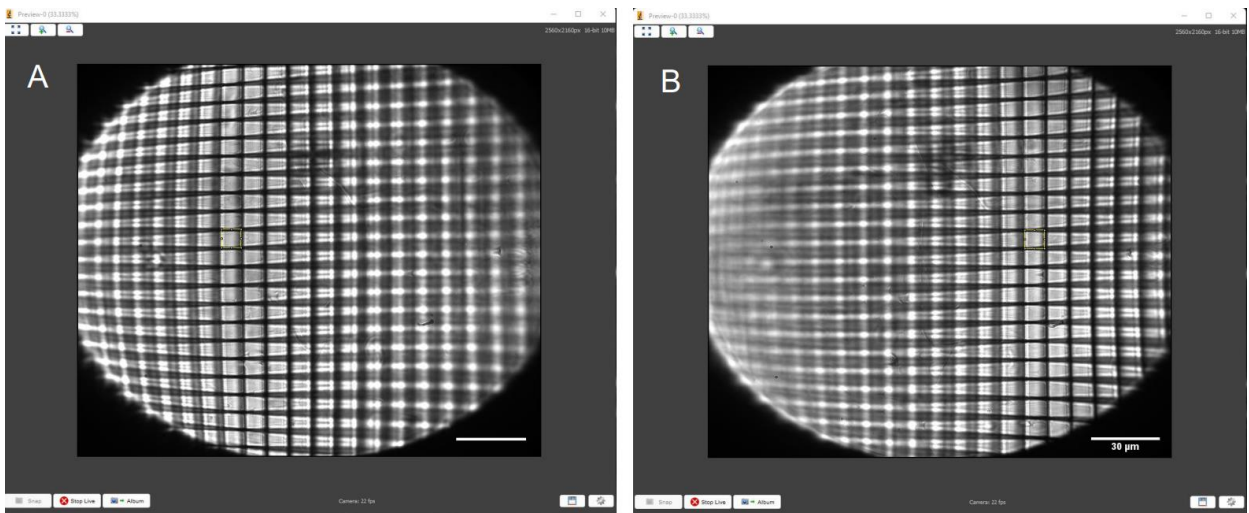

**Figure 10: Positive grid test target with consistent size yellow square overlaid to match the squares of the grid. A.** Grid in focus on the left-hand side. **B.** Grid in focus on the right-hand side. The yellow square matches the size of the grid boxes on both sides of the FoV.

4. **Align the oblique light-sheet:**

4.1) Remove O1 and re-insert the double frosted glass alignment cage in its place. The beam should be collimated and centered on both frosted glass disks.

4.2) Screw CL1 into a rotating lens mount. Mount CL1 into the optical path and rotate the mount so that the beam is expanded in the direction vertical to the optical table. Adjust the tilt and lateral position of CL1 so that the beam is centered on the front and maintains a centered position on both frosted glass disks.

4.3) Screw CL2 into a rotating lens mount and mount CL2 into the optical path at the correct distance to form a 4f system with CL1. Rotate CL2 to the same orientation as CL1 so that the beam is stretched in the direction vertical to the optical table and collimated. Use a test card to measure the height of the cylindrical beam profile at multiple locations to ensure that the beam is collimated. Adjust the tilt and lateral position of CL2 as performed in step 4.2.

4.4) Insert the slit: Mount the slit in a vertical orientation at the front focal plane of L3, measured with a ruler. Use the stretched excitation beam profile to adjust the height and lateral position of the slit until the beam is centered.

4.5) Screw CL3 into a rotating lens mount and mount CL3 into the optical path to form a 4f system with L3 with a shared focal point at the slit. The distance between CL2 and CL3 is not important because the beam is collimated in this space. Rotate CL3 to the same orientation as both CL1 and CL2 so that the beam is focused down to a horizontal sheet profile at the slit. Adjust the tilt and lateral position of CL3 as performed in step 4.2.

4.6) Reinsert O1, mount the fluorescent dye test sample and illuminate the sample with the excitation light-sheet. At the camera sensor the 0° light-sheet should appear as a thin vertical sheet **(Figure 11A).**

4.7) Remove the fluorescent dye test sample and wipe O1 clean. Let the light-sheet expand above O1 unobstructed. Using the motorized translation stage control, translate M1 **towards** the cylindrical lenses to set the angle of the light-sheet to roughly 60° relative to the optical axis of O1.

   NOTE: It Is crucial that the light-sheet is tilted in the correct direction to align with the similarly tilted imaging plane **(Figure 12)**; if a system is laid out different from this specific design, the correct direction of tilt can be figured out through geometric ray tracing.

   NOTE: For reference, translating M1 2.647mm towards the slit set the light-sheet to the correct tilt in this setup.

4.8) Reinsert the fluorescent dye test sample to image the tilted sheet**.** The light-sheet should have maintained a vertical beam shape on camera, but should be wider and fainter **(Figure 11B).**

4.9) Translate the sample axially with the stage so that the fluorescent dye is illuminated by the light-sheet at different depths. Use Fiji to measure the displacement of the center of the light-sheet from the center of the FoV.

4.10) Plot the displacement of the light-sheet as a function of sample depth to calculate the angle of the light-sheet above O1.

4.11) Translate M1 slightly. Repeat steps 4.9 and 4.10 until the angle of the light-sheet is 60° from the optical axis of O1, matching the angle of the imaging plane.

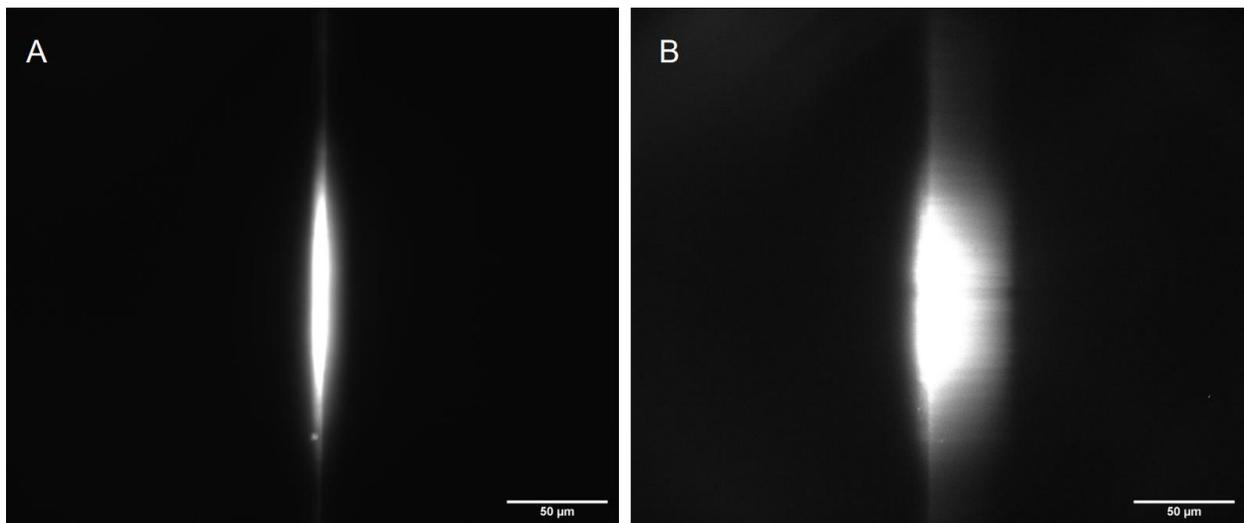

**Figure 11: Camera images of fluorescent dye test sample illuminated by correctly shaped light-sheet at A.** 90°, straight up along the optical axis of O1, and **B.** tilted to 30° (60° to the optical axis of O1).

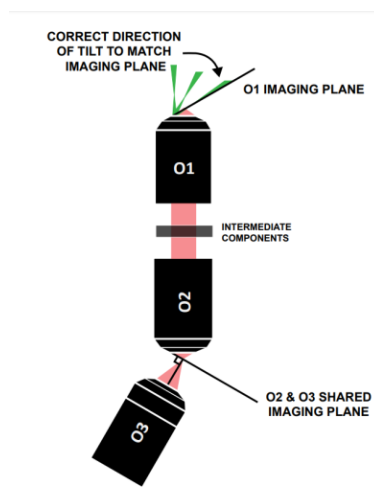

**Figure 12:** Correct direction of light-sheet tilt to align with the imaging plane of O1.

5. **Fine tune the system for imaging and data collection:**

5.1) Mount the 3D bead slide and translate the sample axially with the stage until the beads fill the FoV on camera.

5.2) Adjust O3 using the xy stage and the cage translation stage, aiming to minimize aberrations and optimize signal to noise ratio in the image **(Figure 13).**

5.3) Adjust the correction collar of O1, aiming to minimize aberrations and optimize signal to noise ratio in the image.

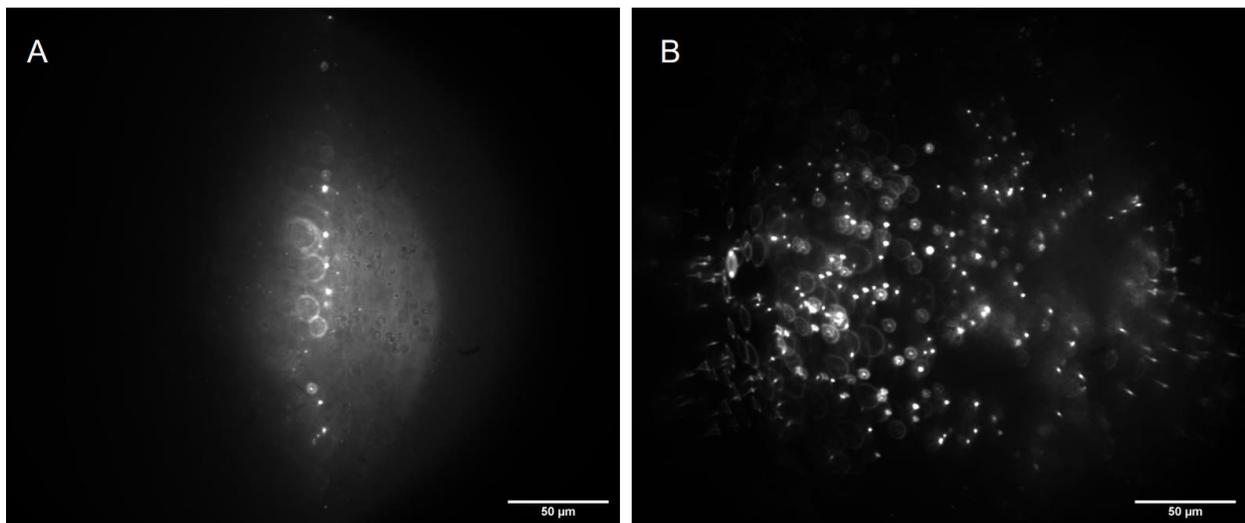

**Figure 13: Camera images of 3D bead sample (1µm beads) illuminated by correctly shaped light-sheet at A.** 90°, straight up along the optical axis of O1, and **B.** tilted to 30° to the optical axis of O1.

6. **Calibrate the magnification of system:**

6.1) Mount the positive grid test target on the sample stage and illuminate with the brightfield light.

6.2) Translate the grid slide axially with the stage to bring the grid slide into focus. Bring the center of the grid into focus.

6.3) Capture and save the image, then open it in Fiji.

6.4) Use the "draw line" and "line profile" functions in Fiji to measure the distance of between two grid lines in pixels. Divide this value by the known distance (10 microns), to determine the pixel to micron calibration.

6.5) Compute the magnification (M) of the system using the measured size of the pixel and the known size of a pixel with:

$$M = \frac{P_{known}}{P_{measured}}$$

7. **Acquiring volumetric scans:**

7.1) Position the sample

7.1.1) Turn on camera preview, galvo, function generator, power supply, stage, and excitation laser.

7.1.2) Mount the sample and set the function generator to triangle wave, then set to following settings to find the sample: 400 mV peak-to-peak amplitude, 0 offset, 100 ms exposure time, 200 mHz frequency.

7.1.3) Scroll in z manually until the sample plane is reached. Optimize the z setting so that the desired region for the volumetric scan passes through the screen during one cycle.

7.2) Select the scan parameter

7.2.1) Ensure that the peak-to-peak amplitude is set correctly by visually checking that the preview looks to be in focus for the duration of the scan. If the image quality degrades sooner approaching one end of the scan than the other, edit the offset on the function generator to move the center of the scan towards the better region.

7.2.2) Select exposure time, framerate, acquisition time, and function generator frequency to best acquire the data. A full volume scan will be created by one half the period of the triangle wave function (linear scan in one direction). If the framerate and frequency are too low, too few frames will be acquired for a volume scan and the low frame number will create visible artifacts in the post-processing. For reference, Figure 13 was composed of ~100 frames in the scan, and Figure 14 from ~800. It is also critical to consider the sample itself in selecting the parameters. Ensure that exposure time is set so that the sample is sufficiently excited, but not saturated. The excitation laser intensity can also be adjusted to this end. If the user is acquiring a series of volumetric scans to characterize a time-varying process in 3D, ensure that the scanning timescale exceeds the timescale of dynamics of the system.

7.3) Collecting videos: Acquire a time-lapse capturing at least the duration of a full ramp up or ramp down of the triangle wave, corresponding to one full scan of the volume.

8. **Post processing procedures:**

8.1) Deskewing volumetric image stacks

8.1.1) The goal of deskewing volume scans is to convert the stack of images in tilted planes to a series of images in real xyz coordinates. There exist many excellent guides on light-sheet image post-processing and open-source software to perform deskewing of existing volume scan s as well as perform and save deskewed images during acquisition[24].

8.1.2) The two parameters required to deskew volume scans are real distance between two frames in pixels (d) and the angle between the frame plane and the x-y plane (\theta). \theta is set by the angle of the oblique light-sheet (in this system, 30°). Distance between frames will depend on the imaging optics and the acquisition settings.

8.2) Finding the d parameter

8.2.1) The distance between frames should be calibrated for every time the system is substantially realigned. We recommend performing this calibration with a stack of images of fluorescent beads, as it is easiest to use to diagnose issues.

8.2.2) Acquire a stack of images and run the deskewing code using any initial guess for the d parameter. Open the deskewed image stack in ImageJ and scroll through the stack. If d has been set substantially far from its true value, the beads will appear artificially elongated in x or y, and individual beads will appear to move in the xy plane as the user scrolls through frames in z (rather than focus and defocus from the same central point). Iterate over the d parameter multiple times until these issues are no longer apparent.

8.2.3) Once the d parameter appears to be reasonably close to the true value, compute max intensity projections of the image stack along the x and y directions. Beads of a diameter near the diffraction limit may appear elongated in z, but ideally should not appear conical or to be elongated diagonally. Fine-tune deskewing parameters until these criteria do not substantially improve with new iterations. For reference, the data shown in figure 13 was deskewed at d = 2.50 pixels and the data in figure 14 was deskewed at d = 1.0 pixels.

NOTE: The distance between frames will depend linearly on scan amplitude, frequency, and framerate.

## REPRESENTATIVE RESULTS:

We perform volumetric scans of 1μm beads embedded in gelrite. Figure 14 shows max intensity projections of deskewed volumetric scans along x, y, and z directions.

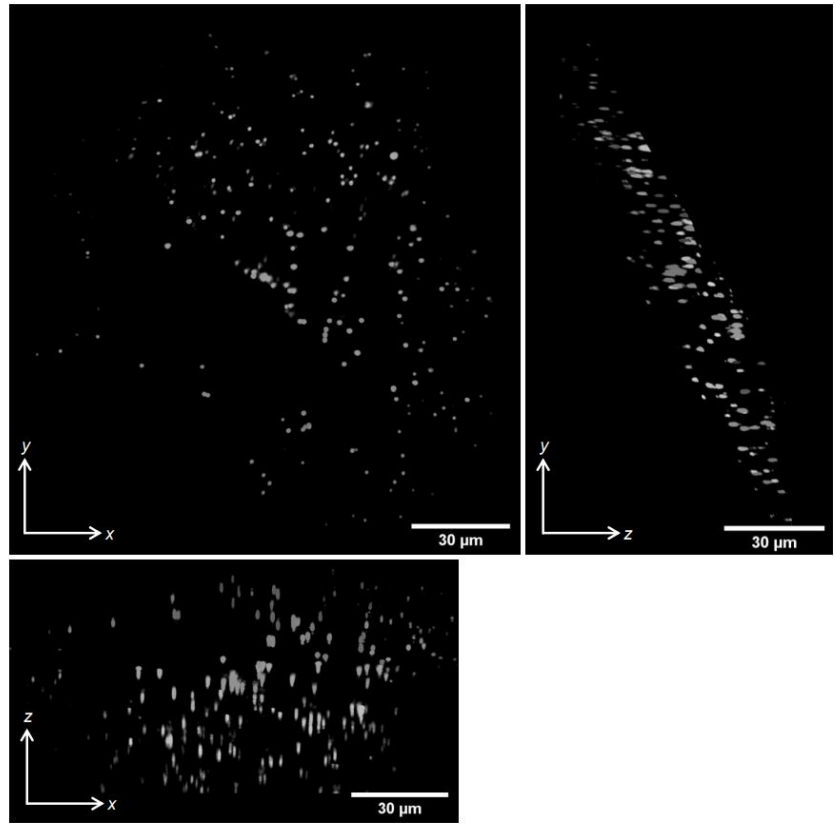

**Figure 14:** Volumetric imaging of 1μm fluorescent beads in gelrite. Max intensity projections of deskewed volumetric scans are shown. Scalebar is 30 μm.

We demonstrate the use of the single-objective light-sheet microscope to characterize reconstituted cytoskeleton networks by performing volumetric scans of samples of microtubule asters. In brief, rhodamine-labeled taxol-stabilized microtubules are polymerized from reconstituted dimers by GTP, then following polymerization, streptavidin-based kinesin motor clusters are mixed into samples along with ATP for final concentrations of 6 μM microtubules, 0.5 μM kinesin dimers, and 10 mM ATP. Extensive protocols and guides for the preparation of taxol-stabilized microtubules and kinesin motor clusters can be found on the Mitchson Lab and Dogic Lab websites[25, 26]. Samples are pipetted gently into microscope slides, sealed, and allowed to sit for eight hours before imaging to allow for motor activity to cease so that samples reach a steady structural state that resembles asters.

Studies of reconstituted cytoskeleton systems most frequently employ confocal or epifluorescence microscopy to image labeled filaments. However, both of these techniques are limited in their capability to image dense 3D samples[27]. While much progress has been made in *in vitro* cytoskeleton-based active matter research by constraining samples to be quasi 2D[28, 29], cytoskeleton networks are inherently 3D, and many current endeavors lie in understanding the effects that can only arise in 3D samples[29, 30], creating a need for high-resolution 3D imaging.

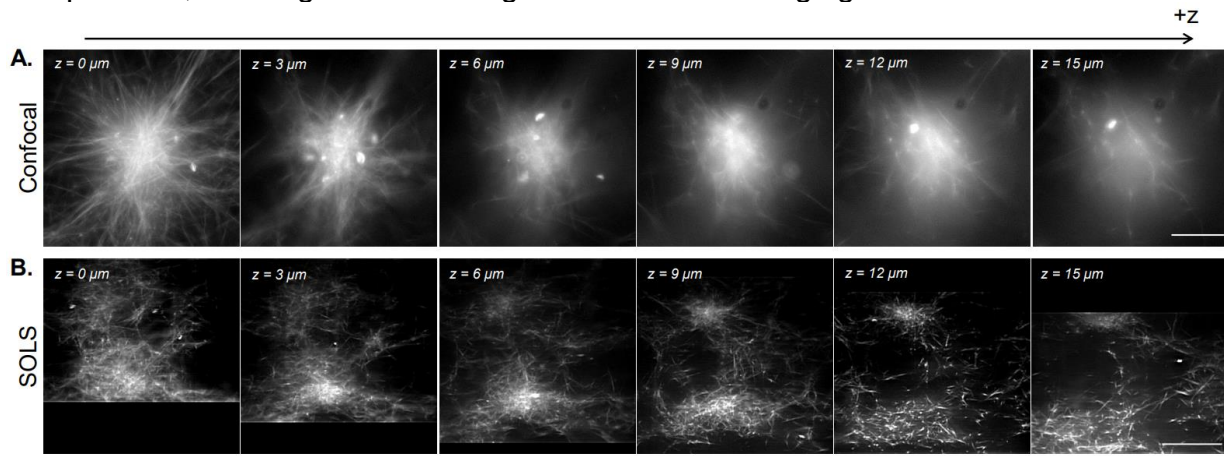

**Figure 15: Single-objective light-sheet microscopy facilitates 3D visualization of reconstituted cytoskeleton samples. A.** Images of fluorescent microtubule asters acquired on a Leica DMi8 laser-scanning confocal microscope. Images show different planes from a z-scan. Scale bar is 30 μm. **B.** Deconvolved deskewed images from a volumetric scan performed on the single-objective light-sheet setup of the same sample. Scale bar is 30 μm. While the confocal excels at imaging single planes near the coverslip, the density of the fluorescent sample introduces complications to imaging at higher planes due to additional signal from below the imaging plane. The light-sheet circumvents this issue by only illuminating the imaging plane, allowing for uniformly sharp imaging at different planes in z.

In **Figure 15** we demonstrate volumetric imaging of a reconstituted microtubule network contracted into aster-like structures by kinesin motor clusters. As shown in previous research [28, 31], these three-dimensional structures tend to grow dense towards the center, resulting in bright regions of fluorescence which predominate the signal. In imaging planes near the coverslip (low z level), confocal microscopy **(Figure 15A)** is able to resolve single filaments around the periphery of the aster, with additional background towards the center due to out-of-focus fluorescence signal from above. However, moving a few microns in z quickly reduces the quality of the images, due to the out-of-focus dense sections of the aster predominating the signal in the imaging plane. The single-plane illumination of the light-sheet **(Figure 15 B)** eliminates the out-of-focus signal from dense parts of the aster above and below the imaging plane, allowing for comparable image quality between planes. The light-sheet's ability to produce high-quality, reliable volumetric scan data opens the door to the possibility of visualizing and characterizing 3D phenomena in reconstituted cytoskeleton systems.

**DISCUSSION:**

There are two critical steps in this protocol. First, the placement of the galvo determines the placement of all other lenses as it is part of three separate 4f lens pairs. It is crucial that the galvo is both conjugated with the back focal planes of both O1 and O2 and centered correctly to ensure tilt-invariant scanning. Second, image quality is extremely sensitive to the alignment of O2 and O3 with respect to each other. Here, care must be taken to ensure that:

1. The alignment angle of O3 to O2 matches the tilt of the excitation light-sheet, providing maximally flat illumination across the similarly tilted FoV
2. O3 is placed at the correct axial distance to maintain a flat FoV with as large of an area as possible.
3. O3 is placed at the correct lateral distance from O2 to maximize the signal that passes through the O2-O3 interface.

We note that this system was not optimized in terms of its total NA, which allows room for significant improvement if users desire a resolving power higher than what this system achieved. There are a multitude of compatible objective options for this type of SOLS build[23], many of which would contribute to a higher resolution of the system with the drawbacks of a higher cost, a smaller FoV or more complicated alignment techniques at the relay interface[8, 11, 13, 20]. Separately, should users desire a larger FoV, incorporating a second galvo to allow 2D scanning would achieve this goal, but would require additional optics and control mechanics to be integrated into the design[32].

Beyond improving the specific components for this particular design, it would be very feasible to add in other high-resolution microscopy techniques or modalities to this build. One such improvement would be to incorporate multi-wavelength illumination, which is a matter of aligning additional excitation lasers to the original excitation path[8]. Furthermore, because this type of SOLS design leaves the sample accessible, adding additional functions to the microscope including but limited to optical tweezing, microfluidics and rheometry is relatively straightforward[2, 33].

Compared to the myriad light-sheet guides that have been published, this protocol provides instructions at a level of understanding that a user without significant optics experience may find helpful. By making a user-friendly SOLS build with traditional sample slide mounting capabilities accessible to a larger audience, we hope to enable an even further expansion of the applications of SOLS based research in all fields of research in which the instrument has or could be utilized. Even with the applications of SOLS instruments rapidly growing in number[2, 34, 35], we believe that many benefits and utilizations of SOLS type instruments still remain unexplored and express excitement at the possibilities for this type of instrument moving forward.

**ACKNOWLEDGMENTS:**
This work was supported the National Science Foundation (NSF) RUI Award (DMR-2203791) to JS. We are grateful for the guidance provided by Dr. Bin Yang and Dr. Manish Kumar during the alignment process, respectively. We thank Dr. Jenny Ross and K. Alice Lindsay for the preparation instructions for the kinesin motors.

**DISCLOSURES:**
The authors have nothing to disclose. All research was conducted in the absence of commercial or financial relationships that could be interpreted as a conflict of interest.